\documentclass[11pt]{JHEP3}
\usepackage{amsmath}
\usepackage{amssymb}
\usepackage{bbold}
\usepackage{stmaryrd}
\usepackage[bottom]{footmisc}

\newcommand{\eqn}[1]{(\ref{#1})}  
\newcommand{\ospsixtwo}{\text{osp}^*(8|4)}

\title{On non-perturbative extensions of anti-de-Sitter algebras}
\preprint{hep-th/0302198\\AEI-2003-026\\SISSA~14/2003/EP}
\author{Patrick Meessen\\
International School for Advanced Studies, \\ 
Via Beirut 2-4,\\ 34014
Trieste, ITALY\\
E-mail: \email{meessen@sissa.it}}
\author{Kasper Peeters\\
MPI/AEI f\"ur Gravitationsphysik\\
Am M\"uhlenberg 1\\
14476 Golm, GERMANY\\
E-mail: \email{kasper.peeters@aei.mpg.de}}
\author{Marija Zamaklar\\
The Abdus Salam ICTP, \\ 
Strada Costiera 11, \\ 
34014 Trieste, ITALY\\
E-mail: \email{mzama@ictp.trieste.it}}

\abstract{Motivated by the study of branes in curved backgrounds, we
  investigate the construction of non-perturbative extensions of the
  super-isometry algebra $\ospsixtwo$ of the $\text{AdS}_7\times S^4$
  background of \mbox{M-theory}. This algebra is not a subalgebra of
  $\text{osp}(1|32)$ and its non-perturbative extension can therefore
  not be obtained by embedding in this simple superalgebra. We show
  how, instead, it is possible to construct an extension directly by
  solving the Jacobi identities. This requires, in addition to the
  expected non-perturbative charges, the introduction of new charges
  which appear in the $\{Q,Q \}$ bracket only via a linear combination
  with the bosonic generators of the isometry algebra.  The resulting
  extended algebra has the correct flat-space limit, but it is not
  simple and the non-perturbative charges do not commute with the
  super-isometry generators.  We comment on the consequences of this
  structure for the representation theory and on possible alternatives
  to our construction.}

\begin{document}
\section{Introduction}

Non-perturbative charges in supersymmetry algebras are very important
for our understanding of \mbox{M-theory}, since they are related to
the presence of branes and the construction of BPS bounds for these
states, which guarantee their stability. Whereas the central extension
of the flat-space super-Poincar\'e algebra has been understood for a
long time and used in many applications, a similar extension for the
super-isometry algebras of the other supersymmetric vacua has not been
used so far and, as a matter of fact, has not yet been constructed
explicitly.\footnote{We prefer to refer to these additional charges in
the algebra as ``non-perturbative'' as opposed to ``central'' in order
to avoid confusion: the charges in this letter will be far from
central.}

There is a widely expressed expectation that the simple algebra
$\text{osp}(1|32)$ plays the role of a ``universal'' algebra for
M-theory, in the sense that the non-perturbatively extended isometry
algebras of the supersymmetric vacua can be obtained as
subalgebras. To be more precise, it is known that the maximal central
extension of the super-translation algebra of flat space can be
obtained as a contraction of the $\text{osp}(1|32)$ algebra, but the
Lorentz generators have to be added by hand.\footnote{The superalgebra
``expansion'' procedure of Azc\'arraga~{\em
et~al.}~\cite{deAzcarraga:2002xi} provides an alternative way to
incorporate the Lorentz generators.}  This construction however, uses
the semi-direct sum structure present in the flat-space algebra: the
bracket of two supertranslation generators never produces a Lorentz
generator. This structure is not present in the isometry algebras of
the other vacua, where the $\{Q,Q\}$ bracket closes into all bosonic
generators. A~quick glance at the literature shows that indeed,
problems arise. For instance, the isometry superalgebra of the
$\text{AdS}_7\times S^4$ vacuum, which is usually denoted as
$\ospsixtwo$, is \emph{not} contained in
$\text{osp}(1|32)$.\footnote{One of the reasons why this subtlety is
often not observed is that the \emph{bosonic} isometry algebra of the
$\text{AdS}_7\times S^4$ background does occur as a subalgebra of
$\text{osp}(1|32)$; see Bars {\em et~al.}~\cite{Bars:1999nk} and
Bandos {\em et~al.}~\cite{Bandos:2002nn} for more on this issue. This
fact will actually be important for us later
(section~\ref{s:reptheory}).}

There is a large body of work based on the~$\ospsixtwo$ algebra. For
instance, it has been used to determine the supergravity Kaluza-Klein
spectrum~\cite{Gunaydin:1985wc}, it is a crucial ingredient in the
construction of the world-volume action for the
supermembrane~\cite{deWit:1998yu} in the associated background and it
plays an important role in checks of the conjectured
eleven-dimensional version of the AdS/CFT
correspondence~\cite{Minwalla:1998rp}. This algebra is thus likely to
be present, in some form or another, in the full algebra that
incorporates perturbative as well as non-perturbative physics. Given
the comments in the previous paragraph, one may be tempted to
conjecture that $\ospsixtwo$ is in fact all there is, and that it
already contains the information about non-perturbative
states~\cite{Ferrara:2000ki}. This is, in our opinion, unlikely to be
correct. One reason is that such a structure is not compatible with a
flat space-time limit, in which central charges are known to
appear. Moreover, it is known that the space-time superalgebra of the
supermembrane in the $\text{AdS}_7\times S^4$ background has to
contain, apart from the $\ospsixtwo$ generators, non-perturbative
charges~\cite{Furuuchi:1999tn} corresponding to winding or infinitely
extending membranes. This suggests that a larger algebra can indeed be
constructed.

In the present letter we will consider extensions of
$\ospsixtwo$ obtained by adding new bosonic generators yet
keeping the number of fermionic generators fixed (less conservative
options are possible and will be discussed elsewhere; see also the
discussion section). Moreover, we will keep the commutators of the
existing bosonic generators fixed. This restriction is motivated by
the fact that, as we have argued above, the isometry algebra is
reproduced by supermembrane charges even when central charges are
taken into account. Because the anti-commutators of the fermions will
get modified, the original algebra cannot be a sub-superalgebra of the
extended one. When extending the $\{Q,Q\}$ bracket we will therefore
be guided by the requirement that the BPS equations of the old algebra
should be obtainable from the new bracket by setting all new
generators to zero.\footnote{This condition is violated if one tries
to use the $\text{osp}(1|32)$ algebra as a starting point for a
non-perturbative extension, since in this algebra the
$\text{so}(6,2)\times \text{so}(5)$ isometry generators appear in the
$\{Q,Q\}$ bracket with coefficients which are different from the ones
in $\ospsixtwo$. While such an approach has been attempted in
the literature, we regard the BPS relations as essential and will
therefore not follow this route.} This corresponds to the condition
that the extended algebra is also valid in the free theory, {\em i.e.}~that
no phase transition and associated change in the multiplet structure
occurs at zero coupling (a reasonable assumption, since such
non-smooth behaviour at zero coupling would invalidate perturbation
theory altogether).

In the first part of this letter we then show that it is indeed
possible to construct a non-perturbative extension of the $\ospsixtwo$
algebra along these lines, by starting from a very general Ansatz and
solving the Jacobi identities.  A fact which one observes at a rather
early stage in this construction is that additional bosonic charges
are required, charges which are not present in the $\text{osp}(1|32)$
algebra. These appear in the $\{Q,Q\}$ bracket, but only in linear
combination with the bosonic generators of the non-extended isometry
algebra. Without these additional charges it is impossible to satisfy
the Jacobi identities.\footnote{If one does not add such charges, one
is extremely limited in the possible non-perturbative extensions. This
has been analysed some time ago in the context of $N=2,3$ and $4$
superconformal algebras by
Bedding~\cite{Bedding:1984uf,Bedding:1985qh} and appeared more
recently in~\cite{Kamimura:2003rx}; see also
\cite{Gunaydin:1998sw,Gunaydin:1998jc,Bianchi:2002gz}. For instance,
in case of the $\text{AdS}_{5}\times S^{5}$ algebra the result is that
only a ``trivial'' u(1) extension is allowed by the Jacobi
identities~\cite{Kamimura:2003rx}. However, we consider such an
algebra without any other charges undesirable for reasons mentioned
above.} As a result, the maximally extended superalgebra we find is
\emph{not} simple, but instead contains an ideal which is isomorphic
to the \emph{bosonic} isometry algebra of the background. Its
structure is depicted in the diagram below:
\begin{equation}
\label{e:thediagram}
\begin{matrix}
\text{osp}(1|32) & \oplus & \Big( \text{so}(6,2)\oplus \text{so}(5)\Big)\\
{\scriptstyle \{ Q,\hat M^{(8,5)}, Z^{(2|1)}, Z^{(4|2)}\}} & & {\scriptstyle
\{ \hat{W}^{(8,5)} \}}\\
\qquad\qquad\quad\searrow & & \swarrow\qquad\qquad\qquad\\
& \ospsixtwo\\
& \hskip-10em
  \hbox to 0pt{$\scriptstyle \{ Q,\,  M^{(8,5)} = \hat{M}^{(8,5)} + \hat{W}^{(8,5)}\}$\hss}
\end{matrix}
\end{equation}
It is important to note that the isometry algebra of the background
does not correspond to the second term in the sum, but instead is
diagonally embedded.  By taking the second term to be isomorphic to
the bosonic isometry algebra of flat space-time, we show that the same
``diagonalisation'' procedure leads to the centrally extended
super-Poincar\'e algebra. This provides evidence for a complete
M-theory algebra of the form $\text{osp}(1|32)\oplus G$ where $G$
contains at least the bosonic isometry algebras of the M-theory vacua
as subalgebras (the specific option of $\text{osp}(1|32)\oplus
\text{osp}(1|32)$ has been suggested in the literature, though based
on different arguments, see {\em e.g.}~\cite{hora4}).

The algebra obtained in this way should satisfy several physical
requirements. First of all, there should exist a contraction to the
flat, maximally extended super-Poincar\'e algebra (broken to
$\text{iso}(6,1)\oplus\text{so}(4)$). We show that our algebra indeed
satisfies this important requirement. 

Secondly, we should argue that the representation theory of the
extended algebra contains multiplets which correspond to the
multiplets of the original algebra. Despite the fact that the
non-extended algebra cannot be a sub-superalgebra of the extended one
(in which case this proof would be trivial), its structure is hidden
in the larger algebra. This in principle makes it possible that the
relevant physical information present in the non-extended algebra
(like the content of the supergraviton and Kaluza-Klein multiplets and
the BPS relations for them) can be ``lifted'' to the extended algebra.
One simple possibility, which is realised in the case of the
super-Poincar\'e extension, is that there exist representations in
which all new generators act trivially on all states. However, it
turns out this is not case for our algebra. This is essentially due to
the fact that the commutation relations between the supercharges and
the new bosonic generators are \emph{non-trivial} this is why
calling these new charges ``central'' would be even less correct than
in flat space-time, and we instead prefer to call these charges
``non-perturbative'').

Another option could be that, despite the fact that the action of the
new charges is non-trivial on the original~$\ospsixtwo$
multiplets, this action is such that one still preserves the old BPS
relations for these states. This can happen if the expectation values
of the new generators on these states vanish, or conspire and cancel
out from the BPS equations. Because the new charges do not commute
with the other generators, one would probably end up with multiplets
with more states than in the non-extended multiplet, but this is not
necessarily a disastrous problem. In principle these states could
correspond to (bound) multi-particle states of the perturbative
spectrum, generated by application of the new ``non-perturbative''
generators on the old perturbative states.\footnote{These new states
can not, however, correspond to branes, since we expect that branes
decouple from the spectrum in the zero coupling limit. The
perturbative multi-particle states, on the other hand, should be still
present in this limit.} In section~\ref{s:reptheory} we investigate
this possibility, and show that indeed some states in the
supergraviton multiplet might fulfill these requirements. However,
there seem to be no general arguments to prove the existence of the
lift, and in fact it seems unlikely that it exists (a definite proof
would involve the construction of explicit representations of the full
algebra).

We thus conclude that an extension of the super-isometry algebras of
M-theory vacua is possible, but that a physically satisfactory
construction will most likely involve the introduction of yet more
generators, over and above the minimal ones that we have
considered. We conclude this letter with a few comments on such other
alternatives, which are currently under investigation.

\section{M-theory on $\text{AdS}_7\times S^4$ and non-perturbative charges}
\subsection{Constructing an extended $\ospsixtwo$ algebra}

The bosonic isometry algebra of the $\text{AdS}_7\times S^4$
background of M-theory is given by $\text{so}(6,2)\times
\text{so}(5)$. Its super-extension is the superalgebra
$\ospsixtwo$ (in order to denote Lie superalgebras we will
follow the notation used by Van Proeyen~\cite{VanProeyen:1999ni}, see
also Frappat {\em et~al.}~\cite{frap1}). This superalgebra can be written
conveniently using a 13-dimensional notation. With this convention the
algebra has the commutation relations\footnote{\label{f:indexrange}
The $A,B,\ldots$ indices are $\text{so}(6,2)$ indices and $I,J,\ldots$
are $\text{so}(5)$ ones. The eight-dimensional spinors are
anti-Weyl. Note that we take the gamma matrices of these two algebras
to be commuting; this makes the relation to the eleven-dimensional
algebra more complicated but simplifies most of the other calculations
in this paper.}
\begin{equation}
\label{e:trivialstuff}
\begin{aligned}
\big[M^{(8)}_{AB}, M^{(8)}_{CD}\big] &= 
  \delta_{AC} M^{(8)}_{BD}\, , \quad&
\big[M^{(5)}_{IJ}, M^{(5)}_{KL}\big] &= 
  \delta_{IK} M^{(5)}_{JL}\, ,\\
{}\big[M^{(8)}_{AB}, Q\big] &= 
  \frac{1}{8}\big(\Gamma_{AB} Q\big)\, , &
{}\big[M^{(5)}_{IJ}, Q\big] &= 
  \frac{1}{8}\big(\Gamma_{IJ} Q\big)
\end{aligned}
\end{equation}
together with
\begin{equation}
\label{e:osp84QQ}
\big\{Q, Q\big\} = 
     \big(\Gamma^{AB} {\cal C}^{-1}_{(13)}\big)\, M^{(8)}_{AB}
 -2\,\big(\Gamma^{IJ} {\cal C}^{-1}_{(13)}\big)\, M^{(5)}_{IJ}\, .
\end{equation}
Anti-symmetrisation on the appropriate indices on the right-hand
side is understood.

As we have mentioned in the introduction, the superalgebra given above
has been used extensively, and appears for instance in the space-time
supersymmetry algebra of the supermembrane. Keeping track of total
derivative terms, central charge terms have been observed to arise in
the $\{Q,Q\}$ bracket of the superalgebra of the membrane in
$\text{AdS}_7\times S^4$; see the work of
Sato~\cite{Sato:1998yx,Sato:1998yu} and
Furuuchi~{\em et~al.}~\cite{Furuuchi:1999tn}. Just as in flat space one
encounters first of all a two-form charge. One does not see a
five-form charge in the membrane algebra since the only term one can
write down vanishes identically, but this charge is expected to appear
in the matrix model (such charges have indeed been found in the pp-wave
limit, see Sugiyama and Yoshida~\cite{Sugiyama:2002rs} for a membrane
analysis and Hyun and Shin~\cite{Hyun:2002cm} for a matrix model
computation).\footnote{One should not confuse these superalgebras with
the world-volume algebras considered by~Bergshoeff
{\em et~al.}~\cite{Bergshoeff:1998bh}; the algebras considered in that paper
involve world-volume spinors and \emph{are} maximally extended.}  Motivated
by these facts, we thus set out to construct a maximal non-perturbative
extension of~\eqn{e:osp84QQ} and the other commutators.

The supercharges appearing in~\eqn{e:osp84QQ} are in the
32-dimensional representation of the bosonic subalgebra, or more
precisely, in the $(\mathbf{8}_s,\mathbf{4})$. In order to determine
the maximal extension of the $\{Q,Q\}$ bracket, one has to consider
the symmetric tensor product
\begin{equation}
\Big((\mathbf{8}_s,\mathbf{4}) \otimes
(\mathbf{8}_s,\mathbf{4})\Big)_{\text{symm}}
 = (\mathbf{28},\mathbf{1}) \oplus
   (\mathbf{1},\mathbf{10}) \oplus
   (\mathbf{28},\mathbf{5}) \oplus
   (\mathbf{35}^-,\mathbf{10}) \, .
\end{equation}
From the representation theory point-of-view, the \emph{minimal} form
of the maximal extension of the algebra, compatible with the presence
of an $\ospsixtwo$ ``core'', should thus at the very least have a
$\{Q,Q\}$ commutator of the form
\begin{multline}
\label{e:QQwrong}
\big\{Q, Q\big\} = 
     \big(\Gamma^{AB} {\cal C}^{-1}_{(13)}\big)\, M^{(8)}_{AB}
 -2\,\big(\Gamma^{IJ} {\cal C}^{-1}_{(13)}\big)\, M^{(5)}_{IJ}\\[1ex]
 +c\big(\Gamma^{AB}\Gamma^{I} {\cal C}^{-1}_{(13)}\big)\, Z^{(2|1)}_{AB\,I}
 +d\big(\Gamma^{ABCD}\Gamma^{IJ} {\cal C}^{-1}_{(13)}\big)\, Z^{(4|2)\, -}_{ABCD\,IJ}\, .
\end{multline}
The minus sign on the charge $Z^{(4|2)\, - }$ denotes that we have
chosen this charge to be anti-self-dual.  Both the left-hand side and
the right-hand side contain $528$ physical components. We should
stress here that~\eqn{e:QQwrong} does not appear in the commutation
relations of the simple algebra $\text{osp}(1|32)$: with the
normalisation of the isometry generators chosen as
in~\eqn{e:trivialstuff}, the relative coefficient between the
$M^{(8)}$ and $M^{(5)}$ generators in the $\{Q,Q\}$ bracket of this
algebra comes out as~$1$, as opposed to the~$-2$ in~\eqn{e:QQwrong}.

Apart from the fact that the algebra based on the
extension~\eqn{e:QQwrong} eventually turns out to be inconsistent with
the Jacobi identities for any non-zero values of~$c$ and~$d$, there is
also a simpler, counting argument that suggests that this Ansatz
cannot be correct. Namely, upon contracting the generators that appear
in the expressions above to the flat space generators, one knows that
the isometry generators~$M$ of flat space-time disappear from the
$\{Q,Q\}$ bracket, leaving only the translation generators. Therefore,
this contraction of~\eqn{e:QQwrong} will necessarily lead to a
disbalance between the number of components on the left-hand side and
the right-hand side, which one knows does not happen in the maximally
extended super-Poincar\'e algebra.

We therefore introduce additional charges $W^{(8)}_{AB}$ and
$W^{(5)}_{IJ}$ and change the bracket of supersymmetry
charges~\eqn{e:QQwrong} to
\begin{multline}
\label{e:QQcorrect}
\big\{Q, Q\big\} = 
     \big(\Gamma^{AB} {\cal C}^{-1}_{(13)}\big)
         \Big( M^{(8)}_{AB} + \tilde{a} W^{(8)}_{AB}\Big)
  +  \big(\Gamma^{IJ} {\cal C}^{-1}_{(13)}\big)
         \Big( -2 M^{(5)}_{IJ} + \tilde b W^{(5)}_{IJ}\Big) \\[1ex]
 +c\big(\Gamma^{AB}\Gamma^{I} {\cal C}^{-1}_{(13)}\big)\, Z^{(2|1)}_{AB\,I}
 +d\big(\Gamma^{ABCD}\Gamma^{IJ} {\cal C}^{-1}_{(13)}\big)\, Z^{(4|2)\, - }_{ABCD\,IJ}\, ,
\end{multline}
We also make the most general Ansatz for the other brackets which is
compatible with representation theory (these can be found in
equations~\eqn{e:QQansatz}--\eqn{e:standardZZansatz} in the appendix).
In order to fix the coefficients, we have systematically analysed the
Jacobi identities, details of which can be found in the appendix.

The upshot of this analysis is the following. Firstly, one observes
that one actually does not have to go through all of the Jacobi
identities in order to fix the coefficients. Once the structure of the
algebra involving the $\{M^{(8)}, W^{(8)}, M^{(5)}, W^{(5)}\}$
generators is fixed, one can diagonalise this sector. One then finds that
an ideal isomorphic to $\text{so}(6,2)\oplus\text{so}(5)$ is generated
by the linear combinations
\begin{equation}
\hat{W}^{(8)} := \frac{1}{2\tilde{m}-1}\Big(\tilde{m}\, M^{(8)} - W^{(8)}\Big)\quad \text{and}\quad
\hat{W}^{(5)} := \frac{1}{2\tilde{n}-1}\Big(\tilde{n}\, M^{(5)} - W^{(5)}\Big)
\end{equation}
The parameters $\tilde{m}$ and $\tilde{n}$ are related
(see~\eqn{e:mnrel}) and correspond to a single rescaling freedom in
the algebra. Note that this ideal is isomorphic but not identical to
the bosonic isometry algebra of the background. The only non-trivial
commutators involving the $\hat{W}$ charges are in fact (suppressing
indices for brevity)
\begin{equation}
\label{e:idealbrackets}
\big[ \hat{W}^{(8)}, \hat{W}^{(8)} \big] = \hat{W}^{(8)}\,,\quad
\big[ \hat{W}^{(5)}, \hat{W}^{(5)} \big] = \hat{W}^{(5)}\,,
\end{equation}
while other ones vanish (and the $\hat W$ charges in particular map the
supercharges to zero).  The other linear combination that one can
make,
\begin{equation}
\hat{M}^{(8)} := \frac{1}{2\tilde{m}-1}\Big((\tilde{m}-1)\, M^{(8)} + W^{(8)}\Big)\quad \text{and}\quad
\hat{M}^{(5)} := \frac{1}{2\tilde{n}-1}\Big((\tilde{n}-1)\, M^{(5)} + W^{(5)}\Big)
\end{equation}
combines with $Q, Z^{(2|1)}$ and $Z^{(4|2)}$. By computing the
$\{Q,Q\}$ bracket one sees that the $\hat{M}^{(8)}$ and
$\hat{M}^{(5)}$ generators now appear with relative coefficient
one. In fact, these charges turn out to generate the
algebra~$\text{osp}(1|32)$. In terms of the original, unhatted
generators, the algebra is given by~\eqn{e:trivialstuff}, the
supercharge bracket~\eqn{e:QQcorrect} (with fixed values of $c$ and
$d$ in terms of $\tilde m$), as well as the commutation relations
\begin{align}
{}\big[M^{(8)},W^{(8)}\big]  &=  W^{(8)} \,,
&  \big[W^{(8)},W^{(8)}\big] &= \tilde m(\tilde m -1)  M^{(8)} + W^{(8)} \\
{}\big[M^{(5)},W^{(5)}\big]  &=  W^{(5)} \,,
& \big[W^{(5)},W^{(5)}\big]  &= \tilde n(\tilde n -1) M^{(5)} + W^{(5)}  \, .
\end{align}
Our non-perturbative charges are even less ``central'' than in the
super-Poincar\'e case; the brackets with the supercharges mimic the
brackets in~\eqn{e:trivialstuff}:
\begin{equation}
\label{e:WQcomm}
\begin{aligned}
\big[ W^{(8)}, Q \big] &= \tfrac{\tilde m}{8}\, \Gamma Q\, ,&
  \big[ W^{(5)}, Q\big] &= \tfrac{\tilde n}{8}\, \Gamma Q\, ,\\[1ex]
\big[ Z^{(2|1)}, Q\big] &= C\, \Gamma Q\, , &
  \big[ Z^{(4|2)}, Q\big] &= D\, \Gamma Q\, ,
\end{aligned}
\end{equation}
(where $C$ and $D$ are again determined in terms of $\tilde m$). The
appearance of non-trivial commutators like~\eqn{e:WQcomm} is in fact
very general. They appear whenever the non-perturbative charges
transform non-trivially under the action of the isometry generators,
and these isometry generators also appear on the right-hand side of
the $\{Q,Q\}$ bracket. In this case a non-trivial $[Q,Z]$ commutator
is required in order to make the $(Q,Q,Z)$ Jacobi identity work.

We also find that the new charges do not commute among each other:
\begin{equation}
\label{e:MZstuff}
\begin{aligned}
{}\big[M^{(8)}, Z^{(2|1)}\big] = \tilde{m}^{-1} \big[W^{(8)},
  Z^{(2|1)}\big] &= Z^{(2|1)}\,, \\
{}\big[M^{(5)}, Z^{(2|1)}\big] = \tilde{n}^{-1} \big[W^{(5)},
  Z^{(2|1)}\big] &= \tfrac{1}{2}\, Z^{(2|1)}\,, \\
{}\big[M^{(8)}, Z^{(4|2)}\big] = \tilde{m}^{-1} \big[W^{(8)},
  Z^{(4|2)}\big] &= 2\, Z^{(4|2)}\,, \\
{}\big[M^{(5)}, Z^{(4|2)}\big] = \tilde{n}^{-1} \big[W^{(5)},
  Z^{(4|2)}\big] &=  Z^{(4|2)}\,.
\end{aligned}
\end{equation}
The commutators between $Z^{(2|1)}$ and $Z^{(4|2)}$ are similarly
non-trivial but more complicated, see the appendix for details.  By a
rescaling of the fermions one can furthermore eliminate the dependence
on $\tilde{m}$ and $\tilde{n}$.

To summarise, the expressions~\eqn{e:QQcorrect}--\eqn{e:MZstuff}
define the maximally extended super-isometry algebra of the
$\text{AdS}_7\times S^4$ vacuum of M-theory. This algebra has the
structure
\begin{equation}
\label{e:thestructure}
\text{osp}(1|32)\oplus \big( \text{so}(6,2)\oplus
\text{so}(5)\big)\, ,
\end{equation}
but it is very important that the non-extended algebra
$\ospsixtwo$ is embedded in this algebra in a non-trivial way,
as depicted in~\eqn{e:thediagram} given in the introduction. We should
stress that, while it may make sense to write the algebra in a basis
which makes~\eqn{e:thestructure} manifest ({\em i.e.}~using the hatted
generators), this is not the natural basis one would encounter in for
instance the supermembrane algebra. Moreover, it obscures the physical
meaning of the~$M$ generators as isometry generators of the
background.

Note that the Jacobi identities leave \emph{no room} for the inclusion
of a parameter which can smoothly tune the new charges to
zero. However, there is some freedom to completely discard
the~$Z^{(4|2)}$ charge, which may be important for certain
world-volume symmetry algebras.

\subsection{The flat space-time limit}

If our algebra is to be the correct one, it has to pass several
physical consistency checks.  One check is that there is a flat
space-time limit in which the algebra reduces to the centrally
extended, \mbox{super-$\text{iso}(6,1)\times \text{so}(4)$} algebra.
(Obviously, one can never recover the eleven-dimensional
super-Poincar\'e vacuum, since the local Lorentz algebra in the
\mbox{$\text{AdS}_7\times S^4$} space-time is broken from
$\text{so}(10,1)$ to \mbox{$\text{so}(6,1)\oplus \text{so}(4)$}, and
contraction does not change the number of generators).

To take the flat space-time limit of the $AdS_7\times S^4$ algebra
presented in the previous section we first rewrite the algebra in a
manifestly \mbox{$\text{so}(6,1) \oplus \text{so}(4)$} symmetric
form. The isometry generators in the two formulations are related
according to 
\begin{align}
\label{scal1}
M_{\underline{1}a}^{(8)} & \rightarrow R P_a \, , &  
   M_{ab}^{(8)} &\rightarrow M_{ab}^{(8)} \, , & 
   W^{(8)}_{\underline{1}a} & \rightarrow R^{\Sigma} W^{(1)}_{a} \, & 
   W^{(8)}_{ab} & \rightarrow R Z^{(2)}_{ab}  \, \\[1ex]
M_{5i}^{(5)} & \rightarrow R P_i \, , &  
   M_{ij}^{(5)} &\rightarrow M_{ij}^{(5)} \, , & 
   W^{(5)}_{5i} & \rightarrow R^{\Sigma} W^{(1)}_{i} \, , & 
   W^{(5)}_{ij} & \rightarrow R Z^{(2)}_{ij}  \, .
\end{align}
where we have split the indices as $A=(\underline{1},a)$ and
$I=(i,5)$. Here $R$ is the dimensionless radius of the AdS$_7$ factor
({\em i.e.}~the dimensionful radius divided by the Planck length) while the
radius of the $S^4$ equals $R/2$. We will comment below on the value
of the parameter $\Sigma$ when we take the limit to flat
space-time. The $Z^{(2|1)}$ generators split as
\begin{equation}
\label{scal2}
Z^{(2|1)}_{\underline{1}a|i} \rightarrow R Z^{(2)}_{(1,1)} \, , \quad 
Z^{(2|1)}_{\underline{1}a|5} \rightarrow R Z^{(5)}_{(1,4)} \, , \quad
Z^{(2|1)}_{ab|5} \rightarrow R Z^{(5)}_{(2,0)}\equiv R Z^{(5)}_{(5,0)}  \, ,\quad
Z^{(2|1)}_{ab|i} \rightarrow R Z^{(5)}_{(2,3)} \, ,
\end{equation}
while the $Z^{(4|2)}$ generators split and scale as
\begin{equation}
Z^{(4|2)}_{\underline{1}abc|ij} \rightarrow R Z^{(5)}_{(3,2)} \, , \quad 
Z^{(4|2)}_{\underline{1}abc|i5} \rightarrow R Z^{(5)}_{(4,1)} \, .
\end{equation}
We used the notation $Z^{(2)}_{(p,q)}$ and $Z^{(5)}_{(p,q)}$ to denote
a two-form resp.~five-form which transforms as a $(p,q)$-form under
the $\text{so}(6,1) \oplus \text{so}(4)$ algebra.  Counting the
generators, we are left with precisely the number of central charges
expected in the flat space-time algebra, except for the two additional
$W^{(1)}$~charges. However, as we will see, these additional charges
do not pose problems. Finally, the supercharges are redefined as
\begin{equation}
Q \rightarrow \sqrt{R} Q \, .
\end{equation} 

The flat space-time limit is now obtained by taking the
$R\rightarrow\infty$ limit. In order to compare the result with a
direct construction of the algebra in flat space-time, one has to
express the $\text{so}(6,2)\oplus\text{so}(5)$ gamma matrices in terms
of $\text{so}(6,1)\oplus\text{so}(4)$ ones (this can be done by using
$\text{so}(10,1)$ intermediate notation as explained in~\eqn{clif1}
and~\eqn{clif2} in the appendix). One easily finds agreement between
these two results. The $\{ Q, Q \}$ bracket turns into
\begin{equation}
\label{e:QQ-11dbreak}
\begin{aligned}
\{ Q , Q \} &= 
     \big(\tilde{\Gamma}^{a}C^{-1}\big)      P_a 
   + \big(\tilde{\Gamma}^i C^{-1}\big)       P_i 
   + \big(\tilde{\Gamma}^{ab} C^{-1}\big)    Z_{ab}^{(2)} 
   + \big(\tilde{\Gamma}^{ij} C^{-1}\big)    Z_{ij}^{(2)} 
   + \big(\tilde{\Gamma}^{ai} \tilde\Gamma C^{-1}\big)    Z_{ai}^{(2)}\\[1ex]
  &+ \big(\tilde{\Gamma}^{abijk} C^{-1}\big) Z_{abijk}^{(5)} 
   + \big(\tilde{\Gamma}^{aijkl} C^{-1}\big) Z_{aijkl}^{(5)} 
   + \big(\tilde{\Gamma}^{abcde}\tilde\Gamma C^{-1}\big) Z_{abcde}^{(5)} \\[1ex]
  &+ \big(\tilde{\Gamma}^{abcdi} C^{-1}\big) Z_{abcdi}^{(5)} 
   + \big(\tilde{\Gamma}^{abcij} C^{-1}\big) Z_{abcij}^{(5)} \, ,
\end{aligned}
\end{equation}
where $\tilde{\Gamma}^a$ and $\tilde\Gamma^{i}$ are gamma matrices in
$C(6,1)$ and $C(4,0)$ respectively, $\tilde\Gamma$ denotes the
chirality matrix in $C(4,0)$ and the charge conjugation matrix is
given by $C \equiv C_{(6,1)} \times C_{(4)}$. Note that the
four-dimensional spinors are not Weyl, which explains the appearance
of~$\tilde\Gamma$, yet dualisation has been freely used in the
seven-dimensional sector (for instance to
identify~$Z^{(5)}_{(2,0)}\equiv Z^{(5)}_{(5,0)}$ in~\eqn{scal2}). The
bracket~\eqn{e:QQ-11dbreak} is precisely what one would get if one
were to construct this flat-space central extension from scratch.

Several comments are in order here.  Firstly, note that flat space
five-form $Z^{(5)}$ arises both from the $Z^{(2|1)}$ and the $Z^{(2|4)}$
charges.  Secondly, the parameter $\Sigma$ in~\eqn{scal1} with which
we scale the $W^{(8)}$ and $W^{5}$ charges should lie in the range
\begin{equation}
\tfrac{1}{2} < \Sigma \leq 1 \, .
\end{equation}
For $\tfrac{1}{2}<\Sigma<1$, the charges $W^{(1)}_a$ and $W^{(1)}_i$
decouple from the remainder of the algebra in the flat space-time
limit: they disappear from the $\{Q,Q\}$ bracket and become mutually
commuting, while keeping their transformation property under the
action of the $M$ generators.\footnote{It might be that this one-form
charge is related to the possibility of constructing kappa-symmetric
open membranes which end on a string~\cite{deWit:1997fp}.} In the case
$\Sigma=1$, these two one-form charges combine with the momenta, and
appear only through the combinations $(P_a+W_a)$ and $(P_i+W_i)$ in
the full algebra. The other linear combination has trivial commutators
with all other generators, including the Lorentz generators $M$. This
case is perhaps more natural and resembles a similar situation in
string theory, where winding charges combine with
momenta~\cite{town6}.

Finally, note that it was crucial for the correct flat space limit
that the generators $M^{8,5}$ and $W^{8,5}$ get scaled away from the
$\{Q , Q \}$ commutator, but \emph{not} $\hat{M}^{(8,5)}$,
$\hat{W}^{(8,5)}$. The latter option would lead to a reduction of the
number of generators on the right hand side of this commutator, while
the left hand side would still contain the same number of
components. Hence, to take the the correct flat space limit it is
crucial to correctly identify the bosonic isometry generator as
belonging to the \emph{diagonal} $\text{so}(6,2) \times \text{so}(4)$
within the algebra~\eqn{e:thestructure}.

\subsection{The super-Poincar\'e algebra from a diagonal embedding}

While the presence of $\text{osp}(1|32)$ in our algebra is reminiscent
of the structure of the maximally extended super-Poincar\'e algebra,
the isometry generators are embedded in a different way, namely
through ``diagonal embedding'' instead of through a semi-direct
sum. In the present section we would like to comment on this
difference and show how the diagonal embedding approach also works for
the flat space-time situation.

Remember that a contraction of the $\text{osp}(1|32)$ algebra leads to
the maximally extended eleven-dimensional super-\emph{translation}
algebra~\cite{town3}, spanned by the translation generators~$P$ and
two- and five-form central charges~$Z^{(2)}$ and~$Z^{(5)}$.  To obtain
the full super-Poincar\'e algebra, one has to add the Lorentz
generators of the $\text{so}(10,1)$ algebra separately:~one starts
with the semi-direct sum $\text{so}(10,1) \inplus \text{osp}(1|32)$
(where $\text{so}(10,1)$ acts non-trivially on $\text{osp}(1|32)$),
and does not scale these generators in the process of contraction (see
however~\cite{deAzcarraga:2002xi}).

Alternatively, inspired by the structure that has appeared in the
previous section, one could start from the algebra
$\text{osp}(1|32)\oplus \text{so}(10,1)$ (where ``$\oplus$'' denotes
the \emph{direct} sum). Let us denote the generators of
$\text{so}(10,1)$ in $\text{osp}(1|32)$ with $\hat{Z}^{(2)}$ and the ones
sitting in the other $\text{so}(10,1)$ with $\hat{M}^{(2)}$.  Decomposing
the $\{Q, Q \}$ commutator with respect to the $\text{so}(10,1)$
subalgebra of $\text{osp}(1|32)$, one gets the familiar expression
\begin{equation} 
{} \{ Q , \, Q \} \sim \hat{P} + \hat{Z}^{(2)} + \hat{Z}^{(5)} \, \, . 
\end{equation}
The rest of the $\text{osp}(1|32)$ commutators can be written schematically as
\begin{align}
\label{flat-comm}
{} &[ \hat{Z}^{(2)} , \, \hat{Z}^{(2)} ] \,  \sim \hat{Z}^{(2)} \quad
\, &  &[ \hat{Z}^{(2)} \,
, \hat{Z}^{(5)} \, ]  \sim \hat{Z}^{(2)} + \hat{Z}^{(5)} \, & 
 & [ \hat{Z}^{(5)}\, , \hat{Z}^{(5)} ] \,  \sim \hat{Z}^{(2)} \, + \, \hat{Z}^{(5)} \\
{} &[ \hat{P}, \hat{Z}^{(2)} ]  \sim \hat{P} \, &  & [ \hat{P}, \hat{Z}^{(5)} ]  \sim \hat{Z}^{(5)} \, & & [
\hat{P}, \hat{P} ]  \sim \hat{P} + \hat{Z}^{(2)} \, .
\end{align}
As in the AdS case, one identifies the Lorentz generators of the flat
vacuum with a diagonal subalgebra of the two~$\text{so}(10,1)$ algebras,
\begin{equation}
M^{(2)} \equiv \hat{M}^{(2)} + \hat{Z}^{(2)} \, , \quad Z^{(2)}
 \equiv \hat{M}^{(2)} - \hat{Z}^{(2)}\,,\quad P \equiv \hat{P}  \, .
\end{equation}
In taking the contraction to the flat space limit, one scales the
generators as
\begin{equation}
\label{scal}
Q \rightarrow \sqrt{R} Q \, , \quad P \rightarrow R P \, , \quad
M \rightarrow M \, ,  \quad Z^{(2)} \rightarrow R
Z^{(2)} \, , \quad Z^{(5)} \rightarrow R Z^{(5)} \, .
\end{equation}
It is now obvious that by rewriting~\eqn{flat-comm} in terms of the
new variables~$M^{(2)}$ and~$Z^{(2)}$ and applying the
scaling~\eqn{scal}, one arrives at the commutation relations of the
full super-Poincar\'e algebra (with trivial commutation relations
between all central charges and a nontrivial action of the Lorentz
generators~$M^{(2)}$ on all central charges).  Note that before the
contraction, the $\{ Q , Q \}$ commutator contains, just as in the AdS
case, a linear combination of the bosonic generators and central
charges. Additionally, the bosonic generators drop out from the
commutator only after having taken the~$R \rightarrow \infty$ limit.

From the two explicit examples we have worked out in detail in this
note, it is suggestive that the general form of the complete
\mbox{M-theory} algebra could be of the form $\text{osp}(1|32) \oplus
G$, where $G$ is a minimal \emph{bosonic} algebra that incorporates
the isometry algebras of all M-theory vacua. However, we still have to
analyse the representation theory of our algebra and compare it with
the representation theory of~$\ospsixtwo$.

\subsection{Representation theory and the graviton multiplet}
\label{s:reptheory}

If the non-perturbatively extended algebra presented in this
paper is to be compatible with existing literature, we should show (at
the very least) that the massless graviton multiplet of the
non-extended~$\ospsixtwo$ algebra ``lifts'' to a representation of the
bigger algebra. The fact that the $\ospsixtwo$ algebra is not a
sub-superalgebra of the algebra we have constructed means that, apart
from the fact that we have introduced new
generators~(non-perturbative charges), we have also modified
the action of the ``old'' generators~(supercharges). A ``lift'' of
existing supermultiplets of $\ospsixtwo$ can therefore not be just a
simple embedding.  Since the new generators of the extended algebra
have non-trivial commutation relations with the supercharges, the
various states in a supermultiplet will in principle transform
non-trivially under the action of the new charges. Hence, a single
state of the old algebra is generically lifted to a set of states with
the same values of the old quantum numbers, yet with non-zero values
of the new charges.

The graviton multiplet is a short multiplet: when constructing the
multiplet by repeatedly acting on the lowest energy states with
energy-raising supersymmetry operators, one encounters some states
which have zero norm and are therefore absent from the
multiplet. Consider now the eigenvalue problem
\begin{equation}
\det \Big( \langle \phi| \{Q_\alpha, Q_\beta\} 
   - \lambda \delta_{\alpha\beta}|\phi\rangle \Big) = 0
\end{equation}
for a given state $|\phi\rangle$. Because of the existence of null
vectors some supercharges act trivially on $|\phi\rangle$, and the
polynomial on the left-hand side will have a number of zeroes at
$\lambda=0$. Some of these are trivial: for {\em e.g.}~a ground state, where
we know that all energy-lowering operators should vanish identically,
we find a factor $\lambda^{16}$.  The non-trivial factor(s) of this
polynomial constitute the BPS equation(s) for the given state
$|\phi\rangle$.

The unitary irreducible representations of $\ospsixtwo$ are
uniquely labeled~\cite{Gunaydin:1985wc} by representations of its
maximal compact bosonic subalgebra
$\text{so}(2)\times\text{su}(4)\times \text{so}(5)$.\footnote{Note
that each unitary irrep of the compact subgroup $\text{so}(2)\times\text{su}(4)$
gives rise to an \emph{infinite-dimensional} representation of the
full non-compact group~$\text{so}(6,2)$. This reflects the general
fact that there are no finite-dimensional unitary representations of
non-compact groups.}  The BPS equations on a state are expressed in
terms of the associated quantum numbers. By ``lifted state'' we now
mean a state in a multiplet of the extended algebra which transforms
in the same way under the bosonic generators inherited from the old
algebra, and is annihilated by the same number of supercharges as
before\footnote{Lifting the representation of the diagonal
subalgebra to the full one can be understood more easily by
considering a toy example, for instance the lift of $\text{su}(2)$
representations to $\text{su}(2)\times \text{su}(2)$ in which it is
embedded.}.  The BPS condition for the state in the extended algebra can
be derived from
\begin{equation}
\langle \phi | \{ Q, Q\} |\phi\rangle = 
\langle \phi | \Big( M^{(8)} - 2 M^{(5)} + W^{(8)} + W^{(5)} + Z^{(2|1)} +
Z^{(4|2)}\Big) | \phi\rangle\, .
\end{equation}
The first two terms on the right-hand side are the same as in the
$\ospsixtwo$ algebra, and hence vanish, when evaluated on any
$\ospsixtwo$ BPS state. Using simple representation theory arguments
one can show that the expectation values of \emph{some} of the
non-perturbative charges in \emph{some} of the gravity
multiplet states vanish. There are, for instance, several
$\text{AdS}_7$ scalars in the graviton
multiplet~\cite{Gunaydin:1985wc} (the ``dilaton'') which carry the
$\text{so}(2)\times\text{su}(4)\times \text{so}(5)$ quantum numbers
$[4,0,0,0,2,0]$. The tensor product with~$Z^{(2|1)}$ does not contain
this dilaton representation, and the presence of
$\langle\phi|Z^{(2|1)}|\phi\rangle$ does therefore not modify the BPS
equation for these particular states.

However, this simple argument cannot be used to deduce the absence of
all non-perturbative charges from the BPS equation for all
states. This is essentially due to the fact that there are charges
which transform in the same way as either $M^{(8)}$ or $M^{(5)}$ under
the action of the $\text{so}(5)$ group. This then implies that their
expectation values in the supergraviton states are generically
non-vanishing. The only way they could drop out from the BPS equations
is if they accidentally vanish for the given representation, or if
they conspire together to give zero.  It seems unlikely that this will
happen in general, but a definite conclusion will require the explicit
construction of the representations of the full algebra, which we will
not attempt here.

\section{Discussion}

We have constructed a maximal non-perturbative extension of the
super-isometry algebra of the \mbox{$\text{AdS}_7\times S^4$} vacuum
of \mbox{M-theory} by adding new bosonic charges to the isometry
algebra \mbox{$\ospsixtwo$} (a similar construction is possible for
the other supersymmetric vacua of \mbox{M-theory} and those of string
theory). These additional charges are even less central than those
that appear in the extension of the super-Poincar\'e algebra: just as
in the supersymmetric extension of conformal algebras~\cite{holt4}
they commute neither with the other bosonic charges nor with the
supercharges. While our algebra satisfies a couple of physical
consistency checks, we have also argued that the structure might lead
to a representation theory which does not produce the expected
``lifted'' $\ospsixtwo$ multiplets.

The extension constructed here exhibits additional non-abelian
charges, which is a generic feature of extensions of AdS
algebras. This implies that a direct interpretation of the new
generators in terms of brane charges becomes more complicated than in
flat space-time. One particular situation in which the analysis may be
simplified is in the Hpp-wave limit of the background discussed in
this paper. Branes in this background have been studied using probe
branes, boundary states and supergravity solutions, but a simple
classification using representation theory of an extended algebra
would be desirable; we intend to investigate this problem in the near
future.

It is conceivable that our extension is still too conservative, and
that the problems in finding a correct perturbative spectrum arise
because an even larger algebra should be considered. Our analysis
shows that one is then \emph{forced} to introduce additional fermionic
charges, which may perhaps lead to an $\text{osp}(1|32)\oplus
\text{osp}(1|32)$ structure. Fermionic extensions actually seem
to be suggested by an analysis of the superalgebra of the matrix model
in a pp-wave background, and we will report on this issue
elsewhere~\cite{mpz2}.

\acknowledgments

We would like to thank Hermann Nicolai, Martin O'Loughlin, Tom\'as
Ort\'\i{}n, Jan Plefka, Andrew Waldron and especially Stefano Kovacs
for discussions.

\vfill\eject

\appendix
\section{Appendix}
\subsection{Conventions and Clifford algebras}

Our $\text{so}(6,2)$ spinors are taken to be anti-Weyl, so together
with our conventions for the $\text{so}(5)$ part we have
\begin{equation}
\Gamma^{{A}_1\cdots {A}_8} = 
   -\epsilon^{{A}_1\cdots {A}_8}\, ,\quad
\Gamma^{{I}_1\cdots {I}_5} =
   \epsilon^{{I}_1\cdots{I}_5}\, .
\end{equation}
This is consistent with our choice for anti-self-duality of the charge
$Z^{(4|2)}$,
\begin{equation}
\label{e:antiselfduality}
Z^-_{{A}{B}{C}{D}|{I}{J}} = 
-\tfrac{1}{4!}\epsilon_{{A}{B}{C}{D}}{}^{{E}{F}{G}{H}}
Z^-_{{E}{F}{G}{H} | {I}{J}}\, .
\end{equation} 

We use the notation $C(t,s)$ for the Clifford algebra for 
a Lorentz-like metric with signature~$(t,s)$. The explicit
representation of $C(6,1)\otimes C(4,0)$ by means of $C(10,1)$ is
given by $(a=0,\ldots,6; \, i=7,\ldots,10)$,
\begin{equation}
\begin{aligned}
\label{clif1}
\tilde{\Gamma}^{i} &= \gamma^{i} \,, & 
\tilde{\Gamma}^{a} &= \gamma^{*}\gamma^{a}\,,  \\
\gamma^{*} &= \gamma^{7\ldots 10}\,, & 
C &= {\cal C}_{(11)}\gamma^* \,,
\end{aligned}
\end{equation}
while a representation of $C(6,2)\otimes C(5,0)$ by means of $C(10,1)$
is given by 
\begin{equation}
\label{clif2}
\begin{aligned}
\Gamma^{i} &= \gamma^{i}\otimes 1\,, & 
  \Gamma^{y} &= -i\gamma^{*}\otimes 1\,,\\[1ex]
\Gamma^{a} &= \gamma^{*}\gamma^{a}\otimes\sigma^{1}\,, &
  \Gamma^{\underline{1}} &= -1\otimes \sigma^{2}\,,
\end{aligned}
\end{equation}
where, in the notation of footnote~\ref{f:indexrange}, the index ranges are
$(A=\underline{1},0,\cdots,6;\, I=7,\ldots,10,y)$.  The charge
conjugation matrix of this 13-dimensional representation and the Weyl
projector in the $\text{so}(6,2)$ factor are given by
\begin{equation}
  {\cal C}_{(13)} = {\cal C}_{(11)}\ \gamma^{*}\otimes 1 
  ,\quad\text{and}\quad
\Gamma_{(9)} = 1\otimes\sigma^{3}\,.
\end{equation}

\subsection{Solving the Jacobi identities}

As explained in the main text, we have derived the
non-perturbative extension of the $\ospsixtwo$ algebra by
starting from a general Ansatz and solving the Jacobi identities. The
key steps are presented below. The most important ingredient in our
Ansatz is the bracket
\begin{equation}
\label{e:QQansatz}
{}\{ Q, Q \} = a M^{(8)} + \tilde{a} W^{(8)} + b M^{(5)} 
  + \tilde{b} W^{(5)} + c Z^{(2|1)} + d Z^{(4|2)} \, \, .
\end{equation}
where we have suppressed all indices and gamma matrices for brevity.
We will use the standard $\text{so}(6,2)$ and $\text{so}(5)$ brackets
for the isometry generators. One may actually think of adding central
charges to these~$[M,M]$ commutators, but apart from the reasons
against this which were given in the introduction, one also quickly
finds out that in this case it is impossible to satisfy the $(Q,Q,Z)$
and $(M,M,Z)$ Jacobi identities at the same time.  The most general
form of the commutators involving the generators $M^{(8)}$, $M^{(5)}$
and the new charges $W^{(8)}$, $W^{(5)}$ is
\begin{equation}
\begin{aligned}
{}\big[M^{(8)},W^{(8)}\big] &= \alpha M^{(8)} + \tilde{\alpha} W^{(8)} \,,
&  \big[W^{(8)},W^{(8)}\big] &= \beta M^{(8)} + \tilde{\beta} W^{(8)} \\
{}\big[M^{(5)},W^{(5)}\big] &= \gamma M^{(5)} + \tilde{\gamma} W^{(5)} \,,
& \big[W^{(5)},W^{(5)}\big] &= \delta M^{(5)} + \tilde{\delta} W^{(5)}  \, .
\end{aligned}
\end{equation}
These generators act on the fermionic charges as
\begin{equation}
\begin{aligned}
\big[ M^{(8)}, Q \big] &= A\, \Gamma Q\, , & 
  \big[ M^{(5)}, Q\big] &= B\, \Gamma Q\, ,\\[1ex]
\big[ W^{(8)}, Q \big] &= A_z\, \Gamma Q\, ,&
  \big[ W^{(5)}, Q\big] &= B_z\, \Gamma Q\, ,\\[1ex]
\big[ Z^{(2|1)}, Q\big] &= C\, \Gamma Q\, , &
  \big[ Z^{(4|2)}, Q\big] &= D\, \Gamma Q\, .
\end{aligned}
\end{equation}
What remains is to fix the commutators with the $Z^{(2|1)}$ and
$Z^{(4|2)}$ charges. These are parametrised as
\begin{equation}
\label{Z21}
\begin{aligned}
{}\big[M^{(8)}, Z^{(2|1)}\big] &= m\, Z^{(2|1)}\,,  &  \big[W^{(8)},
Z^{(2|1)}\big] &= \tilde{m} \big[M^{(8)}, Z^{(2|1)}\big] \,, \\
{}\big[M^{(5)}, Z^{(2|1)}\big] &= n\, Z^{(2|1)}\,,  &  \big[W^{(5)},
Z^{(2|1)}\big] &= \tilde{n} \big[M^{(5)}, Z^{(2|1)}\big]\, ,
\end{aligned}
\end{equation}
and
\begin{equation}
\label{Z42}
\begin{aligned}
{}\big[M^{(8)}, Z^{(4|2)}\big] = \mu\, Z^{(4|2)}\,, \quad \quad &  \big[W^{(8)}, Z^{(4|2)}\big] = \tilde{\mu} \big[M^{(8)}, Z^{(4|2)}\big]\,, \\
{}\big[M^{(5)}, Z^{(4|2)}\big] = \nu\, Z^{(4|2)}\,, \quad \quad &  \big[W^{(5)}, Z^{(4|2)}\big] = \tilde{\nu} \big[M^{(5)}, Z^{(4|2)}\big]\,.
\end{aligned}
\end{equation}
The ``standard'' non-perturbative charges obey the commutators
\begin{equation}
\label{e:standardZZansatz}
\begin{aligned}
{}\big[Z^{(2|1)}, Z^{(2|1)}\big] &= p M^{{(8)}} + \tilde{p} Z^{{(8)}} + q M^{4} + \tilde{q} W^{(5)} + r Z^{(4|2)} \, ,\\[1ex]
{}\big[Z^{(2|1)}, Z^{(4|2)}\big] &= 
  \rho \Big( Z^{(2|1)} - \frac{1}{4!} \epsilon_8 Z^{(2|1)}\Big)
  +\sigma \epsilon_5 Z^{(4|2)}\, ,\\[1ex]
{}\big[Z^{(4|2)}, Z^{(4|2)}\big] &=  
   (\delta - \tfrac{2}{4!}\epsilon ) (\lambda M^{(8)} + \tilde\lambda W^{(8)})
 + (\delta - \tfrac{1}{4!}\epsilon ) (\kappa M^{(5)} + \tilde\kappa W^{(5)} )\\[1ex]
 &\quad + \chi (\delta - \tfrac{1}{4!}\epsilon) Z^{(2|1)}
 + \phi Z^{(4|2)}\, .
\end{aligned}
\end{equation}
The structure involving the $\delta$ and $\epsilon$ parts is
determined by imposing self-duality on the tensors on the left-hand
side; explicit indices can easily be inserted.

The parameters $a$ and $b$ are fixed to be $a=1$ and $b=-2$ because we
want the structure of the $\ospsixtwo$ algebra to be manifestly
present in the $\{Q,Q\}$ bracket. In order to determine the remaining
parameters, we first scale the generators $W^{(8)}$, $W^{(5)}$ and
$Z^{(1|1)}$ in such a way that
\begin{equation}
\label{e:scalings}
\tilde\beta = 1\, ,\quad
\tilde\delta = 1\, ,\quad
p = \pm 1\, ,
\end{equation}
which is always possible. The $(M,M,W)$ Jacobi identities restrict the
way in which the two-form charges act on each other; from
$(M^{(8)},M^{(8)},W^{(8)})$ one finds
\begin{equation}
\{ \alpha=0\,,\; \tilde\alpha=1\} \quad\text{or}\quad \tilde\alpha=0\, .
\end{equation}
Similar conditions for $\gamma$ and $\tilde\gamma$ arise from
$(M^{(5)},M^{(5)},W^{(5)})$.  We discard the second option because it
makes the interpretation of the $M$~charges as isometry generators
troublesome. The action of these isometry generators on the
``standard'' non-perturbative charges $Z^{(2|1)}$ and $Z^{(4|2)}$ is fixed by
imposing the following Jacobi identities:
\begin{equation}
\begin{aligned}
\big(M^{(8)},M^{(8)},Z^{(2|1)}\big) &\rightarrow m=1\, ,&
\big(M^{(5)},M^{(5)},Z^{(2|1)}\big) &\rightarrow n=\tfrac{1}{2}\, ,\\[1ex]
\big(M^{(8)},M^{(8)},Z^{(4|2)}\big) &\rightarrow \mu=2\, ,&
\big(M^{(5)},M^{(5)},Z^{(4|2)}\big) &\rightarrow \nu=1\, .
\end{aligned}
\end{equation}
One now first fixes the action of all two-form charges on the fermions,
using for instance the Jacobi identities
\begin{equation}
\begin{aligned}
\big(M^{(8)}, Q, Q\big) &\rightarrow A =\tfrac{1}{8}\, ,&
\big(M^{(5)}, Q, Q\big) &\rightarrow B =\tfrac{1}{8}\, ,\\[1ex]
\big(W^{(8)}, Q, Q\big) &\rightarrow A_z =\tfrac{\tilde m}{8}=\tfrac{\tilde\mu}{8}\, ,&
\big(W^{(5)}, Q, Q\big) &\rightarrow B_z =\tfrac{\tilde n}{8}=\tfrac{\tilde\nu}{8}\, .
\end{aligned}
\end{equation}
The structure of the algebra in the sector of the two-form charges is
then fixed to a large extent from the Jacobi identities
\begin{equation}
\begin{aligned}
\big(W^{(8)}, W^{(8)}, Q\big) &\rightarrow \beta = \tilde m(\tilde m
-1)\, ,\\[1ex]
\big(W^{(5)}, W^{(5)}, Q\big) &\rightarrow \delta = \tilde n(\tilde n
-1 )\, ,
\end{aligned}
\end{equation}
and using a different part of the $(Z,Q,Q)$ identities,
\begin{equation}
\begin{aligned}
\big(W^{(8)}, Q, Q\big) &\rightarrow \tilde a = \frac{1}{\tilde m
-1}\, ,\\[1ex]
\big(W^{(5)}, Q, Q\big) &\rightarrow \tilde b = \frac{-2}{\tilde n
-1}\, .
\end{aligned}
\end{equation}
A key relation between the two remaining parameters $\tilde m$ and
$\tilde n$ is found from
\begin{equation}
\label{e:mnrel}
\begin{aligned}
\big( Z^{(2|1)}, Q, Q \big) &\rightarrow \frac{2\tilde m-1}{\tilde
m-1} = 8\,cC = -2\frac{2\tilde n-1}{\tilde n-1}\, ,\\[1ex]
\big( Z^{(4|2)}, Q, Q \big) &\rightarrow cC = 2\cdot 4! dD\, .
\end{aligned}
\end{equation}
This last line ensures that the triple-$Q$ identity is satisfied,
since it leads to
\begin{equation}
\big(Q,Q,Q\big) \rightarrow 2(aA + \tilde{a} A_z) 
+ (bB + \tilde{b}B_z) + 2\,cC - 10\cdot 4!\, dD\, .
\end{equation}
Finally, one determines the way in which the $W$ charges appear in the
$[Z^{(2|1)},Z^{(2|1)}]$ bracket by considering
\begin{equation}
\begin{aligned}
(Z^{(2|1)},Z^{(2|1)}, W^{(8)}) &\rightarrow \tilde p =
\frac{1}{\tilde{m}-1}\, ,\\[1ex]
(Z^{(2|1)},Z^{(2|1)}, W^{(5)}) &\rightarrow \tilde q =
\frac{1}{\tilde{n}-1}\, ,\\[1ex]
(Q,Q,Z^{(2|1)}) &\rightarrow q=\mp 1\, ,
\end{aligned}
\end{equation}
(the sign in the last equation depending on the sign of~$p$
in~\eqn{e:scalings}). Analogously one determines similar 
coefficients in the $[Z^{(4|2)}, Z^{(4|2)}]$ bracket from
\begin{equation}
(Q,Q,Z^{(4|2)}) \rightarrow (\tilde{m}-1) \tilde\lambda = \lambda\,,\quad
                            (\tilde{n}-1) \tilde\kappa  = \kappa\, .
\end{equation}

The Jacobi identities not listed so far do not impose any further
restrictions on the parameters. At this stage, one can now diagonalise
the algebra in the two-form sector, as explained in the main text, and
obtain the structure given in~\eqn{e:thestructure}.

\bibliographystyle{JHEP}
\bibliography{mesa_letter}
\end{document}